# Can magnetic multilayers propel artificial micro-swimmers mimicking sperm cells?

François Alouges*, Antonio DeSimone†, Laetitia Giraldi‡, Marta Zoppello§

**Abstract**

We formulate and solve the equations governing the dynamics of a microscopic artificial swimmer composed of a head and of a tail made of a thin film of permanent magnetic material. This is a variant of the model swimmer proposed by Dreyfus *et al.* in 2005, whose tail is a filament obtained from the assembly of super-paramagnetic beads. The swimmer is actuated by an oscillating magnetic field, and its geometry is inspired by that of sperm cells. Using values for the geometric and material parameters which are realistic for a magnetic multi-layer, we show that the model swimmer can reach swimming speeds exceeding one body-length per second, under reasonable values of the driving magnetic field. This provides a proof of principle for the viability of the concept. In addition, we discuss the possibility to steer the system along curved paths. Finally, we compare the propulsion mechanism (swimming 'gait') of our swimmer with that of sperm cells. The main difference between the two is that, contrary to its biological template, our artificial system does not rely on the propagation of bending waves along the tail, at least for the range of material and geometric parameters explored in this paper.

**Keywords**: low Reynolds number swimmers, scallop theorem, magnetic multi-layers, artificial flagella and cilia, bio-inspired micro-robots.

*Centre de Mathématiques Appliquées, Ecole Polytechnique, Route de Saclay, 91128 Palaiseau Cedex, France, francois.alouges@polytechnique.edu

†**Corresponding author**. SISSA, Via Bonomea 265, 34136 Trieste, Italy, desimone@sissa.it

‡Unité de Mathématiques Appliquées, ENSTA, Route de Saclay, 91128 Palaiseau Cedex, France, laetitia.giraldi@polytechnique.edu

§ Università degli Studi di Padova, Via Trieste 63, 35121 Padova, Italia, mzoppell@math.unipd.it

# Introduction

Cells and unicellular organisms provide striking examples of microscopic self-propelled objects, with length scales in the range from one to one-hundred microns, that are able to move freely inside the human body. In fact, traffic of self-propelled creatures inside our bodies is quite intense: from leukocytes crawling along tissues and rushing towards a newly opened wound in order to start the immune system response, to muscle cells contracting thanks to myosin motor proteins walking along actin filaments, to a number of micro-organisms swimming inside various cavities and lumens. Two of the most widely studied examples of such microscopic swimmers, for which quantitative models and detailed fluid dynamic data are now available, are bacteria and sperm cells, see, e.g.[1-5] and the many references quoted therein.

While the idea of building artificial devices emulating these motile capabilities is quite natural, much remains to be done for this to be practical. Learning skills from biological organisms requires, in particular, that we learn how to move and control continuously deformable objects such as filaments, cilia, and flagella. This is, in fact, an instance of bio-inspired soft robotics, where novel designs are inspired by the study of how animals exploit soft materials to move effectively in complex and unpredictable natural environment [6-11].

Artificial devices mimicking sperm cells, in which the payload could be contained in a relatively large head, and propulsion forces could be extracted from the beating of a long, thin tail, is a natural concept which has been pioneered in Dreyfus et al.[12]. The idea is particularly attractive also because it may lend itself to diverse micro-fabrication techniques. For example, one may consider functionalized magnetic multi-layers (MMLs) originally conceived for spintronics applications, an idea explored in Courcier et al.[13]. The flexibility in the fabrication procedure could be exploited to target diverse biotechnological applications, by including different functional components. In addition, it could be used to tune the magnetic and elastic properties in order to optimize performance, controllability, manoeuvrability. For the time being, attention is mostly focused on actuation with an externally applied magnetic field, but autonomous self-propelled systems can also be envisaged by using built-in motors, such as muscle cells like in Feinberg et al.[14], or by using active materials, such as in Sawa et al.[15].

Having in mind the applications discussed above, in this paper we present and discuss a computational tool for the simulation of the behaviour of a model magneto-elastic swimmer, consisting of a head and of a tail made of a film of permanent magnetic material, and activated by an oscillating magnetic field. This system is inspired by the pioneering concept explored in Dreyfus et al.[12], based on a magnetic filament consisting of super-paramagnetic beads, and thoroughly analyzed [16,17]. Recent work on a simpler system made of two rigid magnetic segments[18] is also relevant to our analysis. Our aim is to provide a feasibility study for the concept of a magneto-elastic swimmer based on MML fabrication techniques.

The main specific questions we address are whether, by restricting oneself to the small parameter window of magnetic, elastic, and geometric parameters that are realistic for MMLs, and to magnetic field amplitude and actuation frequencies achievable in a laboratory, reasonable swimming speeds can be obtained and, if so, thanks to which swimming gaits. Our answers to these questions are based on a simplified model that makes the problems of control and motion planning tractable.

The problem is complex, and inherently multiphysics. Indeed, it combines magnetism, elasticity, and fluid dynamics in a single system where magnetic torques drive the shape changes of an elastic flagellum which, in turn, produce a propulsive force through the interaction with a surrounding viscous fluid. While a full description of the system via three coupled systems of partial differential equations is, in principle, feasible, our aim is here to develop an agile numerical tool that may help the design, optimization, and motion planning stages.

Our simplifying assumptions reduce the governing equations of our magneto-elastic swimmer to a system of ordinary differential equations (ODEs). Solving these does not require the complex three-dimensional meshing necessary for the numerical solution of the coupled system of partial differential equations of elasticity, magnetostatics, and hydrodynamics. By contrast, our system of ODEs can be easily and quickly solved on a small laptop computer. So, exploring the effect of varying geometric, material, and actuation parameters becomes a feasible task. Given the length scales involved, the induced flows are characterized by very low Reynolds numbers. Accordingly, and in view of the slenderness of the tails, we use the local drag approximation of Resistive Force Theory[19]. Bending of the tail is rendered by concentrating the elasticity on a finite number of points, so that the tail is modeled as a sequence of (many) rigid segments joined by angular springs. Finally, the magnetic behavior of the segments is modeled by assuming that their magnetization is always parallel to the segment, and with fixed magnitude, and that stray fields can be neglected. In future work, we will remove some of the simplifying assumptions leading to the reduced model, as this may be required to resolve some finer details.

The main results of our work are the following. First, we show that by actuating a system made of a non-magnetic head and an MML tail with a magnetic field composed of a constant longitudinal component and an oscillatory transversal one, one can propel it along the longitudinal axis achieving swimming speeds comparable to those observed for bull sperm cells in Friedrich et al.[3], and using magnetic fields that are easily attainable in a laboratory. We use for the magnetic film the material parameters of Permalloy, and geometric parameters that are in the range of current manufacturing techniques. This proves that the MML swimmer is a viable concept, at least in principle.

In addition, we compare the swimming gait of our MML swimmer with that of other natural and artificial micro-swimmers and, in particular, with sperm cells, whose behaviour is well known from the existing literature. It turns out that the mechanisms underlying the motility of MML swimmers and of sperm cells are radically different. Sperm cells propel themselves by propagating bending waves along the flagellum. Similarly, the behaviour of the model swimmer in Dreyfus et al.[12] can be understood as arising from the propagation of bending waves between free and tethered ends, and it is shown in Roper et al.[17] that the resulting gait is intermediate between that of eukaryotic sperm cells and the one of a waggled elastic rod. By contrast, our MML swimmer moves in the absence of bending waves, through a mechanism similar to the one propelling the two-link system studied in [18]. The main difference is that, in our case, the two rigid links with an angular elastic joint are replaced by a flexible magnetic tail, which exhibits a time-dependent spatially constant curvature.

Finally, we show that the longitudinal magnetic field can be used as a steering device, and that by varying its direction one can guide the magneto-elastic swimmer along curved trajectories and even bends of sharply curved pipes.

**Formulation of the problem**

Following [20], we think of our swimmer as composed by $N$ segments $(L_i)_{1 \leq i \leq N}$, which move in the plane $z = 0$. The first segment is special, as it describes the non-magnetic 'head' where the payload is located. Accordingly, it experiences different hydrodynamic drag, as described below. The other segments are characterized by thickness $t_i$ (in direction perpendicular to the filament axis), width $w_i$ (in direction perpendicular to the plane $z = 0$), and length $l_i$ (along the filament axis) such that $t_i \ll w_i \ll l_i$. We take $l_1 = l_{head}$, $t_1 = t_{head}$, $w_1 = w_{head}$, and $l_i = l_{tail}$, $w_i = w_{tail}$, $t_i = t_{tail}$ for $i = 2, \dots, N$. The actual values used for these geometric parameters in the concrete examples analyzed later are given in Table 1.

The position at time $t$ of segment $L_i$ is specified by the position of its first end $\boldsymbol{x}_i = (x_i, y_i, 0)$, and the angle $\theta_i$ that $L_i$ makes with the $x$−axis. We also denote by $\boldsymbol{e}_{i,\parallel} = (\cos \theta_i, \sin \theta_i, 0)$ (resp. $\boldsymbol{e}_{i,\perp} = (-\sin \theta_i, \cos \theta_i, 0)$ ) the unit vector along (resp. orthogonal to) the axis of segment $L_i$.

The segments are linked together, so that the first end of $L_{i+1}$ coincides with the second end of $L_i$, namely, $\boldsymbol{x}_{i+1} = \boldsymbol{x}_i + l_i \boldsymbol{e}_i$. We rewrite these kinematic constraints in the explicit form

$$\begin{cases} x_{i+1} = x_i + l_i \cos(\theta_i), \\ y_{i+1} = y_i + l_i \sin(\theta_i) . \end{cases} \quad (1)$$

The three different physical mechanisms governing the motion of our magneto-elastic swimmer are rendered in the way described below.

*Elasticity*

We account for the elasticity of the structure by using a discrete beam theory. At the junction between segments $L_i$ and $L_{i+1}$, a torsional spring with spring constant $\kappa$ independent of $i$ is assumed to be present. The spring exerts a torque with the same magnitude $\boldsymbol{T}^{el}_{i,\boldsymbol{x}_i} = \kappa(\theta_{i+1} - \theta_i)\boldsymbol{e}_z$, but with opposite signs, on each of the two neighboring segments $L_i$ and $L_{i+1}$ of the swimmer. The spring constant is given by

$$\kappa = \frac{(EJ)_{tail}}{l_{tail}} \quad (2)$$

where $E$ is Young's modulus and $J$ is the moment of inertia of the cross-section of the tail segments

$$J = \frac{1}{12} w_{tail} t_{tail}^3.$$

The actual values used in the next section for these geometric and material parameters are given in Table 1.

*Hydrodynamics*

We assume for simplicity that the swimmer is neutrally buoyant, and that its size and the actuation frequency are such that the induced flows are governed by low Reynolds number hydrodynamics[21]. We model the interaction with the surrounding fluid by using the local drag approximation of Resistive Force Theory[19]. This assumes a linear dependence between the hydrodynamic drag force per unit length acting on the swimmer at a point $\boldsymbol{x}$ and the velocity at that point through the relation

$$\boldsymbol{f}^h(\boldsymbol{x}) = -\xi_i \boldsymbol{u}_\parallel(\boldsymbol{x}) - \eta_i \boldsymbol{u}_\perp(\boldsymbol{x}). \quad (3)$$

Here $\boldsymbol{x}$ is the current location of a point on the $i$-th link, while $\boldsymbol{u}_\parallel(\boldsymbol{x})$ and $\boldsymbol{u}_\perp(\boldsymbol{x})$ stand for the components of the velocity vector of the swimmer at $\boldsymbol{x}$ (and thus of the fluid at the same point $\boldsymbol{x}$, due to the no-slip boundary condition) along $\boldsymbol{e}_i$ and $\boldsymbol{e}_{i,\perp}$, respectively.

The shortcomings of the local drag approximation are well known. In particular, the relation (3) being local, hydrodynamic interactions between the different elements of the swimmer are neglected. Nevertheless, it gives satisfactory results, which are often in striking agreement with experiments, at least for very slender filaments in low Reynolds number flows (see e.g. Friedrich et al.[3]).

Noticing that at $\boldsymbol{x} = \boldsymbol{x}_i + s\boldsymbol{e}_{i,\parallel}$, we have $\boldsymbol{u}(\boldsymbol{x}) = \dot{\boldsymbol{x}}_i + s\dot{\theta}_i \boldsymbol{e}_{i,\perp}$, we can compute the total hydrodynamic force on $L_i$ which is given by

$$\boldsymbol{F}_i^h = \int_0^{L_i} \boldsymbol{f}^h(\boldsymbol{x})\, d\boldsymbol{x} = -l_i \xi_i (\dot{\boldsymbol{x}}_i \cdot \boldsymbol{e}_{i,\parallel})\boldsymbol{e}_{i,\parallel} - \left(l_i \eta_i (\dot{\boldsymbol{x}}_i \cdot \boldsymbol{e}_{i,\perp}) + \frac{l_i^2}{2}\eta_i \dot{\theta}_i\right)\boldsymbol{e}_{i,\perp}. \quad (4)$$

Similarly, the (component perpendicular to the plane z = 0 of the) torque with respect to any point $\boldsymbol{x}_0$ is given by

$$\boldsymbol{e}_z \cdot \boldsymbol{T}_{i,\boldsymbol{x}_0}^h = \boldsymbol{e}_z \cdot \int_0^{L_i}(\boldsymbol{x}-\boldsymbol{x}_0)\times \boldsymbol{f}_h(\boldsymbol{x})\,d\boldsymbol{x} = -\frac{l_i^2}{2}\eta_i(\dot{\boldsymbol{x}}_i\cdot \boldsymbol{e}_{i,\perp}) + \frac{l_i^3}{3}\eta_i \dot{\theta}_i$$
$$+ (\boldsymbol{x}_i - \boldsymbol{x}_0)\times \left(l_i \xi_i (\dot{\boldsymbol{x}}_i \cdot \boldsymbol{e}_{i,\parallel})\boldsymbol{e}_{i,\parallel} + \left(l_i \eta_i (\dot{\boldsymbol{x}}_i \cdot \boldsymbol{e}_{i,\perp}) + \frac{l_i^2}{2}\eta_i \dot{\theta}_i\right)\boldsymbol{e}_{i,\perp}\right) \cdot \boldsymbol{e}_z. \quad (5)$$

For simplicity, we will assume that the drag coefficients are constant along the tail and set $\xi_i = \zeta_{tail,\parallel}$, $\eta_i = \zeta_{tail,\perp}$, for $i = 2, \ldots, N$. The first segment describing the 'head' is special, and we take $\xi_1 = \eta_1 = \zeta_{head}$. The actual values used in Section 3 for these material parameters are given in Table 1.

*Magnetism*

We assume that each segment, excluding only the first one describing the head, is constantly magnetized, and we make the simplifying assumption that the magnetization on each segment stays permanently aligned with the segment axis. We also neglect the magneto-static coupling between different segments, in particular through the stray-field induced by the magnetic distribution along the swimmer. The only magnetic interaction we consider is that with an external applied field: we assume that each segment experiences a (magnetic) torque due to the external magnetic field that is imposed to the swimmer. For the $i$−th segment, this torque takes the form

$$\boldsymbol{T}_i^m = \boldsymbol{M}_i \times \boldsymbol{B} \quad (6)$$

where $\boldsymbol{M}_i$ is the (total) magnetization of the $i$−th segment, $\boldsymbol{B}$ the external magnetic field, and $i$ ranges from 2 to $N$. In view of our assumptions, the magnetization of the $i$−th segment can be written as

$$\boldsymbol{M}_i = M\, l_i\, \boldsymbol{e}_{i,\parallel}. \quad (7)$$

Here $M$ stands for the magnetization per unit length of each segment, which is given by

$$M = M_s t_{tail} w_{tail},$$

where $M_s$ is the saturation magnetization. The actual values used, in the section giving the results, for these material parameters are given in Table 1.

*Governing equations*

It remains to assemble the equations governing the motion of the magneto-elastic swimmer, by putting together the various contributions to forces and torques described above.
We start by observing that the swimmer is completely described - both for its position and shape - by the *3N* variables $(x_i, y_i, \theta_i)_{1 \leq i \leq N}$ satisfying the *2(N − 1)* constraints (1). Therefore, we need to write $N + 2$ additional equations. To that aim, and recalling that the motion takes place in the plane *z = 0*, we write the balance of forces on the whole system (2 equations) and a balance of the torque components perpendicular to *z = 0* with respect to $\boldsymbol{x}_k$ on each of the subsystems consisting of all the segments from *k* to *N*, for *k = 1,⋯,N* (*N* equations). Since inertia is assumed to be negligible, and since the (spatially uniform) external magnetic field exerts no forces but only torques on the various parts of the swimmer, these equations take the form

$$\begin{cases} \boldsymbol{F} = \sum_{i=1}^{N} \boldsymbol{F}_i^h = 0, \\ \boldsymbol{e}_z \cdot \sum_{i=1}^{N}(\boldsymbol{T}_{i,\boldsymbol{x}_1}^h + \boldsymbol{T}_i^m) = 0, \\ \boldsymbol{e}_z \cdot \sum_{i=2}^{N}(\boldsymbol{T}_{i,\boldsymbol{x}_2}^h + \boldsymbol{T}_i^m) = -\kappa(\theta_2 - \theta_1), \\ \vdots \\ \boldsymbol{e}_z \cdot \sum_{i=k}^{N}(\boldsymbol{T}_{i,\boldsymbol{x}_k}^h + \boldsymbol{T}_i^m) = -\kappa(\theta_k - \theta_{k-1}), \\ \vdots \\ \boldsymbol{e}_z \cdot (\boldsymbol{T}_{N,\boldsymbol{x}_N}^h + \boldsymbol{T}_N^m) = -\kappa(\theta_N - \theta_{N-1}). \end{cases} \quad (8)$$

In view of equations (4), (5), (6) and (7), we see that all quantities appearing in system (8) above depend linearly on the rate of positional and orientational changes $(\dot{x}_i, \dot{y}_i, \dot{\theta}_i)_{1 \leq i \leq N}$. Therefore, if we append to equations (8) above the time derivative of the $2(N-1)$ constraints (1), namely,

$$\begin{cases} \dot{x}_{i+1} - \dot{x}_i + l_i \sin(\theta_i)\dot{\theta}_i = 0 \\ \dot{y}_{i+1} - \dot{y}_i + l_i \cos(\theta_i)\dot{\theta}_i = 0 \end{cases} \quad (9)$$

then we end up with a system of ODEs which completely determines the evolution of the magneto-elastic swimmer. This system takes the form

$$A \begin{pmatrix} \dot{x}_1 \\ \vdots \\ \dot{x}_N \\ \dot{y}_1 \\ \vdots \\ \dot{y}_N \\ \dot{\theta}_1 \\ \vdots \\ \dot{\theta}_N \end{pmatrix} = \boldsymbol{F}_0 + \boldsymbol{F}_1 B_x(t) + \boldsymbol{F}_2 B_y(t). \quad (10)$$

The explicit expressions of the matrix $A$ and of the vector-fields $\boldsymbol{F}_0$, $\boldsymbol{F}_1$, and $\boldsymbol{F}_2$ are given in Appendix A. Assembling and solving this system numerically, for a given external field $\boldsymbol{B}(t) = (B_x(t), B_y(t))$ is a relatively straightforward task, see e.g. Alouges et al.[20].

## A case study

We consider the swimmer depicted in Figure 1, which consists of a large (say, disk shaped) head linked to a tail composed of 10 segments. Each segment, including the head, is $10\mu m$ long, so that the length of the whole system is $110\mu m$. For the head, we take $w_{head} = l_{head}$ and $t_{head} = t_{tail}$. For the magneto-elastic parameters we use the values of Permalloy: $E = 10^{11}\ Nm^{-2}$ and $M_s = 8 \cdot 10^5\ Am^{-1}$. As for the drag coefficients, we follow Friedrich et al.[3] and take $\zeta_{tail,\parallel} = 6.2 \cdot 10^{-3}\ Nsm^{-2}$, $\zeta_{tail,\perp} = 12.4 \cdot 10^{-3}\ Nsm^{-2}$, $\zeta_{head} = 0.05\ Nsm^{-2}$. The values for the other parameters used in the numerical simulations are given in Table 1.

*Straight swimming*

We first consider the case where the swimmer, originally in the horizontal position, is excited by a magnetic field with a constant horizontal component and an oscillating vertical one

$$\boldsymbol{B}(t) = (B_x, B_y \sin(\omega t))^t \qquad (11)$$

where $B_x$, $B_y$ have the (fixed) values given in Table 1. These values have been selected on a trial-and-error basis, as field strengths of magnitude achievable in a laboratory and producing interesting performance. Notice that the presence of a nonzero value of $B_x$ proved necessary to obtain stable net motion along the horizontal axis.

We explore the dynamics of the swimmer by varying the driving frequency $\omega/2\pi$ in the range 3-70 *Hz*. We see from Figure 2 that the net horizontal displacement per cycle is maximized at about 8 Hz, while the maximal swimming speed is attained around 50 *Hz*. The value of this maximal displacement is close to 5 μm, while the maximal swimming speed is around 70 *μm/s*.

The evolving shape of the swimmer is well characterized by the angle Ψ*(s, t)* between the horizontal axis and the tangent to the swimmer at arc-length distance s from the external end of the head segment. Following [3], we compute the Fourier coefficients of Ψ*(s, ·)*

$$\widehat{\Psi}_n(s) = \int_0^{\frac{2\pi}{\omega}} \Psi(s,t) \exp(in\omega t)\, dt$$

in order to capture its periodic behavior, and remark that only the term $\widehat{\Psi}_1(s)$ corresponding to the smaller frequency (i.e. the frequency of the magnetic field) is non negligible. We plot in Fig. 3 the complex values of $\widehat{\Psi}_1(s)$ normalized in such a way that $\widehat{\Psi}_1(0)$ is real (in other words, we plot $\frac{\widehat{\Psi}_1(s)\,\overline{\widehat{\Psi}_1(0)}}{|\widehat{\Psi}_1(0)|}$ ). These graphs, shown for the three frequencies highlighted in Figure 2, clearly show that $\widehat{\Psi}_1(s)$ is well approximated by a function of the type $\widehat{\Psi}_1(s) = \lambda + \mu s\, \exp(i\varphi)$ which indicates a behaviour of $\Psi(s, t)$ that is well approximated by the function

$$\Psi(s,t) \sim \mathrm{Re}(\widehat{\Psi}_1(s)\exp(i\omega t)) \sim \lambda \cos(\omega t) + \mu s \cos(\omega t + \varphi)\,. \qquad (12)$$

The deformation of the swimmer is thus composed of a global rotation (the spatially constant term) and of a term describing bending with a spatially constant curvature (the term linear in *s*), which both oscillate in time with angular frequency *ω* and a phase shift *φ*. According to (12), there is no travelling wave of curvature propagating along the tail of the swimmer. Therefore, this swimming mechanism is very different from the one observed in sperm cells[3], but also from the one observed in the artificial system described in Dreyfus *et al.*[12], which is also actuated by an external oscillating magnetic field. In particular, notice that by differentiating (12) with respect to *s*, we obtain that the curvature remains constant along the tail of the swimmer (i.e., *s*-independent) at every time, while being modulated by a time-dependent amplitude.

*Swimming in circles*

The previous section shows that, as it was pointed out already in Dreyfus *et al.*[12], the constant horizontal component of the magnetic field (which is parallel to the initial straight configuration of the magnetic tail, and then parallel to its average orientation during the motion), acts in a stabilizing way, keeping the average orientation of the swimmer always aligned with it. Indeed, the swimmer oscillates, following the oscillations of the transversal component of the applied field, but its average motion is that of a translation along the average direction of the oscillating magnetic field, which is horizontal.
If we now consider an external magnetic field which is obtained by superposing fast transversal oscillations with frequency $\omega$ on a slowly varying longitudinal field, oscillating at frequency $\omega' \ll \omega$, we expect that we can use the direction of the slowly varying field to steer the swimmer. As an example, consider an external magnetic field of the form

$$\boldsymbol{B} = B_\parallel \boldsymbol{e}_{\theta(t)} + B_\perp \sin(\omega t) \boldsymbol{e}_{\theta(t)}^\perp \qquad (13)$$

where $\boldsymbol{e}_{\theta(t)}$ is the unit vector forming an angle $\theta(t)$ with the horizontal axis given by

$$\theta(t) = 2\pi t / T_{max} \qquad (14)$$

and $B_\parallel$ and $B_\perp$ have the same values of $B_x$ and $B_y$, respectively, given in Table 1.
Here, in order to have a clear separation of the time scales associated with fast and slow oscillations, we take $\omega/2\pi = 8$ $Hz$, and $T_{max} = 40s$, which, in view of (14) leads to a frequency $\omega'/2\pi = 0.025$ $Hz \ll \omega/2\pi$.
The swimmer traces now a circular trajectory, and its average orientation follows the slow modulations of the applied magnetic field (see Fig. 5, where only the part of the trajectory following one quarter of a circle is shown).

*Turning abruptly*

In this last section we push further the idea developed in the previous section. Indeed, we take the same parameters as before, given in Tab. 1, and use now a magnetic field given by (13) which oscillates around an average orientation $\boldsymbol{e}_{\theta(t)}$ that now varies in time according to

$$\theta(t) = \frac{\pi}{4}\left(1 + \tanh\left(30\left(\frac{t}{T_{max}} - \frac{1}{2}\right)\right)\right). \qquad (15)$$

Notice that $\theta(t)$ experiences a sudden jump from 0 to $\frac{\pi}{2}$ around $t = \frac{T_{max}}{2}$. The result we obtain is displayed in Fig. 6 and shows clearly a sudden change in the swimming direction, which would allow the swimmer to navigate along an elbow in a pipe. Here, we are tacitly assuming that the pipe is wide enough with respect to the size of the swimmer so that the hydrodynamics effects of the walls can be neglected. Enriching the model to consider explicitly the confining effects due to the pipe walls would be interesting, also in view of recent results in [22], but will not be done here.

*The swimming mechanism: propagation of bending waves along the tail is not necessary for propulsion*

In order to shed light on the mechanism propelling our swimmer, it is useful to introduce the angles $\theta_R := \theta_N$ and $\theta_L := \theta_1$ giving the orientations of the right-most segment and of the left-most one (the head), respectively. Figure 7 shows that the dynamics is such that the point $(\theta_L(t), \theta_R(t))$ traces a loop. By contrast, the orientation of the second link is always very near the one of the first link, and the corresponding loop (shown in red in Figure 7) is close to a single line. The right panel shows snapshots of the swimming stroke along the beat cycle and the dots in the left panel locate them along the loop in the $(\theta_L, \theta_R)$ plane.

The presence of a loop in the $(\theta_L, \theta_R)$ plane shows that the dynamics of the swimmer is not time reversible: net motion in the horizontal direction arises precisely from this lack of time reversibility. In order to make this statement clearer and more quantitative, we compare the swimming gait of our swimmer with the one of a simplified system consisting of two rigid magnetic links joined by an elastic spring where all the bending elasticity is concentrated. A somewhat similar system has been analyzed in [18]. In our case, the first link has also a passive head attached, and hence experiences larger hydrodynamic forces and torques than those acting on the second link. The total length and the magnetic properties of the two swimmers are otherwise identical. Figure 8 shows that the larger hydrodynamic forces acting on the left link cause a delay of its response with respect to the right link, hence a loop in the $(\theta_L, \theta_R)$ plane. Thus, the behaviour of the simpler two-link system reproduces the one of our original system, made of a stiff but deformable magnetoelastic tail.

Figure 8 also shows that no loop is generated in the two-link system when the passive head is removed (dashed curve). Indeed, in this case the two links are subject to the same hydrodynamic forces and no net displacement is produced, as expected.

The dynamics of the simpler two-link system can be easily analyzed. Indeed, in this case, the horizontal velocity of the left-most end of the swimmer is given in terms of $(\theta_L, \theta_R)$ as

$$\dot{x} = g_L(\theta_L, \theta_R)\dot{\theta}_L + g_R(\theta_L, \theta_R)\dot{\theta}_R$$

where the functions $g_L$, $g_R$ are defined in Appendix B.

Integrating this equation over a swimming cycle, and using Stokes theorem, we obtain

$$\Delta x = -\iint_\gamma \text{curl}(g_L, g_R)(\theta_L, \theta_R) d\theta_L d\theta_R$$

where $\gamma$ denotes the region enclosed by the closed loop traced by $(\theta_1(t), \theta_2(t))$ during the cycle. The minus sign comes from the fact that the loops are traced clockwise (in the direction ABCD in Figures 7 and 8). Assuming that the amplitudes of the two angles are sufficiently small, the leading order term of $\Delta x$ becomes

$$\Delta x \simeq - \text{Area}(\gamma)\, \text{curl}(g_L, g_R)(0,0)\,. \tag{16}$$

We can therefore conclude that the net horizontal displacement is proportional to the area of the loop, with a non-vanishing factor given by

$$\text{curl}(g_L, g_R)(0,0) = -\frac{5}{2} \frac{l(\zeta_{tail,\perp} - \zeta_{tail,\parallel})(50\zeta_{tail,\perp} + 11\zeta_{head})}{(10\zeta_{tail,\parallel} + \zeta_{head})(10\zeta_{tail,\perp} + \zeta_{head})}$$

(see Appendix B).

## Discussion and Conclusions

The results of our analysis provide a feasibility study for the engineering of microscopic artificial swimmers consisting of a cargo head and of a flexible thin film tail made of a permanent magnetic material, and propelled by an external oscillating magnetic field. Our results indicate that for a system characterized by geometric parameters consistent with those achievable by current manufacturing techniques, and by realistic values of the magneto-elastic parameters (consistent with those of Permalloy), interesting swimming performance can be achieved by using magnetic fields that are easily attainable in a laboratory (field magnitude of a few tens of *mT*, frequencies of a few tens of Hz).

As shown in Figure 2, the maximum displacement per cycle we obtain is 5 $\mu m$, namely, 0.05 body lengths, and the maximum swimming speed is 70 $\mu m/s$ namely, 0.64 body lengths per second. By tuning and optimizing the geometry and magneto-elastic properties of the tail, one can easily obtain better performance. Consider, for example, the case of a tail made of a magnetic multilayer consisting of a magnetic core of thickness $t_1 = 50\ nm$ and Young modulus $E_1 = E$ (here we use again the material parameters of Permalloy given in Table 1), coated by two non-magnetic layers of thickness $t_2$ and Young modulus $E_2$ (we take $t_2 = 25\ nm$ and $E_2 = 70\ GPa$, the value for Al). Then, formula (2) above is replaced by

$$\kappa = \frac{(EJ)_{eff}}{l_{tail}}$$

where

$$(EJ)_{eff} = \frac{w_{tail}}{12}\left(E_1 t_1^3 + 8E_2\left(\left(\frac{t_1}{2}+t_2\right)^3 - \left(\frac{t_1}{2}\right)^3\right)\right)$$

while the magnetization per unit length of each segment reads

$$M = M_s t_1 w_{tail}\,.$$

This more compliant tail leads to an increase of performance, as shown in Figure 9. We obtain a maximum displacement per cycle, which is now 6.5 $\mu m$, and a maximum swimming speed of 90 $\mu m/s$.

A further increase in performance can be obtained with an even more compliant design, in which the non-magnetic coating is removed in a central portion of the tail of length 10 $\mu m$. This central section acts as an additional elastic joint, where large bending deformations are localized. These lead to larger loops in the $(\theta_L, \theta_R)$ plane (not shown), and the maximum displacement per cycle and the maximum speed reach the values 9.5 $\mu m$ and 125 $\mu m/s$, respectively (see Figure 8). This speed is larger than one body length per second, and exceeds the ones observed for bull sperm cells in water, which are reported to be around 100 $\mu m/s$ [3]. We finally remark that a swimmer made with a very flexible tail consisting of only a 50 $nm$ thick layer with the magnetoelastic properties of Permalloy would in principle produce even better performance. But whether a 110 $\mu m$ long MML film that is only 50 $nm$ thick can be realized in practice, maintaining integrity and mechanical stability when actuated, is possibly questionable. Exploring further variations in the design, and optimizing them subject to the constraints of practical realizability and mechanical integrity is obviously a very interesting problem, but this is beyond the scope of the present paper.

Our analysis shows that the magneto-elastic swimmer we have described in this paper propels itself with a mechanism, which is very different from the ones previously reported in the literature for flexible magneto-elastic filaments. Indeed, the deformation of the swimmer is composed of a global rotation and of a bending deformation with a spatially constant curvature, which both oscillate in time at the same frequency of the external magnetic field, but with a phase shift, see (12). A movie illustrating the corresponding motion can be found in [23].

By contrast, sperm cells and artificial swimmers exploiting control of their curvature propel themselves by propagating internally activated waves of bending along the flagellum[3, 20]. This mechanism can be understood in terms of the classical swimming sheet model of G.I. Taylor[24], since the flagellum is able to produce travelling waves of bending, propagating from tail to head. Consistently with the swimming sheet analysis, this produces motion with average speed in direction opposite to the one of the travelling bending waves, hence head first.

The behaviour of the magneto-elastic filament driven by an external oscillating magnetic field presented in Dreyfus et al.[12] has also been analyzed through the same paradigm. Indeed, its behaviour is rationalized in terms of travelling waves of bending which are now emerging from external activation, rather than being internally produced, and are then observed to propagate from tail to head. This swimmer swims tail first, rather than head first, while exhibiting bending waves propagating form tail to head, again in agreement with the swimming sheet paradigm. In the more detailed study contained in Roper et al.[17], it is further shown that filament propulsion arising from the propagation of bending waves between free and tethered ends leads to a swimming gait that is intermediate between a eukaryotic cell and a waggled elastic rod (see, in particular, Fig. 7 in [17]).

By contrast, in our case there are no travelling bending waves, as shown in the discussion of equation (12). Just like Purcell's idealized scallop[25], whose shape is described by the angle between the two rigid links mimicking the valves, the shape of our swimmer is governed by a single scalar parameter, the spatially constant curvature, evolving in time in a reciprocal fashion. Nevertheless, the swimmer exhibits nonzero net displacements under cyclic actuation.

There is, however, no contradiction with the celebrated Scallop Theorem[25], which applies to systems subject to zero external forces and torques, and powered only by periodic shape changes. In the presence of external torques, the picture changes completely, and non-reciprocal translational motion is no longer incompatible with shape changes that are reciprocal. This has been already observed, e.g., in Burton et al.[26], where it is shown that in the presence of external torques due to the offset between center of mass and center of buoyancy, a two-link device (Purcell's idealized scallop) can swim at low Reynolds numbers. Moreover, it has been recently shown in [18] that a magnetic system made with two rigid segments, one of which magnetic, joined by an angular elastic spring can be propelled by magnetic fields of the same type we have considered. Our swimmer is propelled by a similar mechanism, with the only difference that our magnetic element is flexible, and it exhibits a time-dependent (but spatially constant) curvature.

As shown above, magnetic multilayers lead to films that are relatively stiff in bending. The consequence is that a magneto-elastic swimmer whose flexible magnetic tail is made out of a magnetic multilayer film will likely be too stiff to support the propagation of bending waves. Nevertheless, it will still be able to swim, thanks to the different mechanism illustrated above. The interesting design principle one learns in this way is that, in the presence of an external torque (here of magnetic origin), it is not necessary to engineer the flexible tail endowing it with deformation modes that are not reciprocal in time. The interaction between global (head) rotation and a standing wave of (tail) curvature, both oscillating in time at the same frequency and with a phase shift as predicted by eq. (12), is enough to produce translational motion.

Contrary to the behaviour of sperm cells, our swimmer swims tail first. Also the steering mechanism we use to produce curved trajectories differs from the one used by sperm cells. Our swimmer curves by maintaining the alignment between its average orientation and the average orientation of the external magnetic field. Sperm cells (and, similarly, artificial bio-mimetic devices based on internal actuation providing curvature control) can turn by actuating their tails with waves of curvature with non-zero spatial average, producing trajectories whose curvatures are proportional to the average curvature of the tail, see [3, 20].

In summary, we have shown through a theoretical and numerical study, that it should be possible to extract interesting swimming performance from a system composed of a passive cargo head, a magnetoelastic tail made of a magnetic multilayer, and activated by an external oscillating field. We have discussed the role of tail stiffness in determining swimming performance and found that, all other things being equal, more flexible tails lead to higher swimming speeds (see Figures 9,10).

Magnetic multilayers are relatively stiff in bending: too stiff to allow the propagation of bending waves along the tail. However, we found that even though stiff tails only deform as standing waves, they can still be used to power a swimmer, thanks to a mechanism that is different from the prevailing paradigm in the field (namely, that swimming can only arise from shape changes that are not time-reversible).

## Acknowledgments

This work was partially funded by the European Research Council through the Advanced Grant 340685-MicroMotility. We thank B. Dieny and H. Joisten for valuable discussions.

# Appendix A

The $3N \times 3N$ matrix $A$ appearing in equation (10) is given by blocks as

$$A = \begin{pmatrix} F_{x,\dot{x}} & F_{x,\dot{y}} & F_{x,\dot{\theta}} \\ F_{y,\dot{x}} & F_{y,\dot{y}} & F_{y,\dot{\theta}} \\ T_{\dot{x}} & T_{\dot{y}} & T_{\dot{\theta}} \\ C_{x,\dot{x}} & 0 & C_{x,\dot{\theta}} \\ 0 & C_{y,\dot{y}} & C_{y,\dot{\theta}} \end{pmatrix}$$

according to the force, torque and constraint equations (8)-(9). The F matrices are $1 \times N$ row vectors given component-wise by

$$(F_{x,\dot{x}})_{1,i} = -l_i(\xi_i \cos^2 \theta_i + \eta_i \sin^2 \theta_i), \quad (F_{y,\dot{x}})_{1,i} = -l_i(\xi_i - \eta_i) \sin \theta_i \cos \theta_i,$$
$$(F_{x,\dot{y}})_{1,i} = (F_{y,\dot{x}})_{1,i}, \quad (F_{x,\dot{y}})_{1,i} = -l_i(\xi_i - \eta_i) \sin \theta_i \cos \theta_i,$$
$$(F_{x,\dot{\theta}})_{1,i} = \frac{l_i^2}{2} \eta_i \sin \theta_i, \quad (F_{y,\dot{\theta}})_{1,i} = -\frac{l_i^2}{2} \eta_i \cos \theta_i,$$

for $i = 1 \cdots N$. Matrices $T$ are $N \times N$ matrices given by

$$(T_{\dot{x}})_{ij} = \eta_j \frac{l_j^2}{2} \sin \theta_j - (x_j - x_i) l_j (\xi_j - \eta_j) \sin \theta_j \cos \theta_j + (y_j - y_i) l_j (\xi_j \cos^2 \theta_j + \eta_j \sin^2 \theta_j),$$

$$(T_{\dot{x}})_{ij} = \eta_j \frac{l_j^2}{2} \cos \theta_j + (x_j - x_i) l_j (\eta_j \cos^2 \theta_j + \xi_j \sin^2 \theta_j) - (y_j - y_i) l_j l_j (\xi_j - \eta_j) \sin \theta_j \cos \theta_j$$

$$(T_{\dot{\theta}})_{ij} = -\eta_j \frac{l_j^3}{3} - (x_j - x_i) \frac{l_j^2}{2} \eta_j \cos \theta_j - (y_j - y_i) \frac{l_j^2}{2} \eta_j \sin \theta_j.$$

for $i, j$ ranging from 1 to $N$. Finally, matrices $C$ are $(N - 1) \times N$ matrices for which the non-vanishing terms are given by

$$(C_{x,\dot{x}})_{ii} = -1, \quad (C_{x,\dot{x}})_{i,i+1} = 1, \quad (C_{x,\dot{\theta}})_{ii} = l_i \sin \theta_i,$$
$$(C_{y,\dot{y}})_{ii} = -1, \quad (C_{y,\dot{y}})_{i,i+1} = 1, \quad (C_{x,\dot{\theta}})_{ii} = -l_i \cos \theta_i,$$

for $i$ ranging from 1 to $N - 1$. For what concerns the vector-fields appearing in equation (12), we have

$$\boldsymbol{F}_0 = -\kappa(0,0,0,\theta_2 - \theta_1, \ldots, \theta_N - \theta_{N-1}, 0, \ldots, 0)^t,$$

$$\boldsymbol{F}_1 = -M_s(0,0, \sum_{i=1}^{N} \sin \theta_i, \ldots, \sum_{i=k}^{N} \sin \theta_i, \ldots, \sin \theta_N, 0, \ldots, 0)^t,$$

$$\boldsymbol{F}_2 = -M_s(0,0, \sum_{i=1}^{N} \cos \theta_i, \ldots, \sum_{i=k}^{N} \cos \theta_i, \ldots, \cos \theta_N, 0, \ldots, 0)^t.$$

Notice that the $2(N - 1)$ zeros at the end of the vector-fields correspond to the (differential) constraint equations (11).

# Appendix B

In Section 3.4 we focus on a swimmer made with two unequal rigid and magnetized links, one of which has a passive head attached, and joined by an elastic rotational spring. The length of each link is $5\,l$, the one of the head is $l$. The drag coefficient of the head is $\zeta_{head}$ and the ones for the tail are $\zeta_{tail,\parallel}$ and $\zeta_{tail,\perp}$.

We derive the equation of motion for this system from (10) by setting $N=3$ and $\theta_2 = \theta_1$. The last assumption is made to fix the orientation of head to be equal to the one of the first segment.

Considering the subsystem made by the first 2 rows of (10) and by noticing that the 2 first components of vector fields $\mathbf{F}_0$, $\mathbf{F}_1$ and $\mathbf{F}_2$ are null (see Appendix A), we get

$$\mathbf{M}\begin{pmatrix}\dot{x}_1\\ \dot{y}_1\end{pmatrix} = (\mathbf{G}_L, \mathbf{G}_R)\begin{pmatrix}\dot{\theta}_L\\ \dot{\theta}_R\end{pmatrix} \qquad (17)$$

where $\mathbf{M}$ is a symmetric matrix defined by

$$\mathbf{M} = \begin{pmatrix} m_{11} & m_{12} \\ m_{12} & m_{22} \end{pmatrix}$$

with

$$m_{11} = 5l\big(\zeta_{tail,\perp} - \zeta_{tail,\parallel}\big)(\cos^2\theta_L + \cos^2\theta_R) - 10\,\zeta_{tail,\perp}\,l - \zeta_{head}\,l,$$
$$m_{12} = 5l\big(\zeta_{tail,\perp} - \zeta_{tail,\parallel}\big)(\sin\theta_L \cos\theta_L + \sin\theta_R \cos\theta_R),$$
$$m_{22} = -5l\big(\zeta_{tail,\perp} - \zeta_{tail,\parallel}\big)(\cos^2\theta_L + \cos^2\theta_R) - 10\,\zeta_{tail,\parallel}\,l - \zeta_{head}\,l.$$

Here $\mathbf{G}_i = (G_i^1, G_i^2)$ with $i = L, R,$ are two column vectors whose expression is

$$G_L^1 = -30l^2 \cos\theta_R\big(\zeta_{tail,\perp} - \zeta_{tail,\parallel}\big)(\sin\theta_L \cos\theta_R - \sin\theta_R \cos\theta_L) + \frac{1}{2}l^2 \sin\theta_L\big(95\zeta_{tail,\perp} + \zeta_{head}\big),$$

$$G_L^2 = -30l^2 \cos\theta_R\big(\zeta_{tail,\perp} - \zeta_{tail,\parallel}\big)(\sin\theta_L \sin\theta_R + \cos\theta_R \cos\theta_L) - \frac{1}{2}l^2 \cos\theta_L\big(\zeta_{head} + 35\zeta_{tail,\perp} + 60\zeta_{tail}$$

$$G_R^1 = \frac{25}{2}\zeta_{tail,\perp} \sin\theta_R\, l^2,$$

$$G_R^2 = -\frac{25}{2}\zeta_{tail,\perp} \cos\theta_R\, l^2.$$

The first component of system (17) is

$$\dot{x} = g_L(\theta_L, \theta_R)\dot{\theta}_L + g_R(\theta_L, \theta_R)\dot{\theta}_R$$

where each function $g_i$ depends on $(\theta_L, \theta_R)$ and is equal to $\mathbf{M}^{-1}\mathbf{G}_i \cdot \mathbf{e}_x$. Integrating this over a cycle we obtain eq. (16), namely,

$$\Delta x \simeq -\text{Area}(\gamma)\text{curl}(g_L, g_R)(0,0).$$

To compute explicitly the quantities above, let us recall that $\text{curl}(g_L, g_R)(\theta_L, \theta_R)$ is defined by

$$\text{curl}(g_L, g_R)(\theta_L, \theta_R) = \frac{\partial g_R}{\partial \theta_L}(\theta_L, \theta_R) - \frac{\partial g_L}{\partial \theta_R}(\theta_L, \theta_R)$$

Therefore, the derivative of $g_i$ with respect to $\theta_j$, $i,j = L, R$ and $i \neq j$, evaluated at $(0,0)$ is equal to

$$\frac{\partial g_i}{\partial \theta_j}(0,0) = \left(\frac{\partial \boldsymbol{M}^{-1}}{\partial \theta_j}\boldsymbol{G}_i + \boldsymbol{M}^{-1}\frac{\partial \boldsymbol{G}_i}{\partial \theta_j}\right)\bigg|_{(\theta_L,\theta_R)=(0,0)} \cdot \boldsymbol{e}_x.$$

Moreover,

$$\frac{\partial \boldsymbol{M}^{-1}}{\partial \theta_j}\bigg|_{(\theta_L,\theta_R)=(0,0)} = -\left(\boldsymbol{M}^{-1}\cdot\frac{\partial \boldsymbol{M}}{\partial \theta_j}\cdot \boldsymbol{M}^{-1}\right)\bigg|_{(\theta_L,\theta_R)=(0,0)}$$

with

$$\boldsymbol{M}^{-1}|_{(\theta_L,\theta_R)=(0,0)} = \frac{1}{l}\begin{pmatrix} \dfrac{1}{-10\zeta_{tail,\parallel} - \zeta_{head}} & 0 \\ 0 & \dfrac{1}{-10\zeta_{tail,\perp} - \zeta_{head}} \end{pmatrix},$$

$$\frac{\partial \boldsymbol{M}}{\partial \theta_j}\bigg|_{(\theta_L,\theta_R)=(0,0)} = 5l\begin{pmatrix} 0 & \zeta_{tail,\perp} - \zeta_{tail,\parallel} \\ \zeta_{tail,\perp} - \zeta_{tail,\parallel} & 0 \end{pmatrix},$$

$$\boldsymbol{G}_L|_{(\theta_L,\theta_R)=(0,0)} = -\frac{l^2}{2}\begin{pmatrix} 0 \\ 95\zeta_{tail,\perp}+\zeta_{head} \end{pmatrix},\quad \boldsymbol{G}_R|_{(\theta_L,\theta_R)=(0,0)} = -\frac{25}{2}l^2\begin{pmatrix} 0 \\ \zeta_{tail,\perp} \end{pmatrix},$$

$$\frac{\partial \boldsymbol{G}_L}{\partial \theta_R}\bigg|_{(\theta_L,\theta_R)=(0,0)} = 30l^2(\zeta_{tail,\perp} - \zeta_{tail,\parallel})\begin{pmatrix}1\\0\end{pmatrix},\quad \frac{\partial \boldsymbol{G}_R}{\partial \theta_L}\bigg|_{(\theta_L,\theta_R)=(0,0)} = \begin{pmatrix}0\\0\end{pmatrix},$$

From these expressions, we get

$$\frac{\partial g_R}{\partial \theta_L}(0,0) = -\frac{125}{2}\frac{l(\zeta_{tail,\perp} - \zeta_{tail,\parallel})\zeta_{tail,\perp}}{(10\zeta_{tail,\parallel} + \zeta_{head})(10\zeta_{tail,\perp} + \zeta_{head})},$$

and

$$\frac{\partial g_L}{\partial \theta_R}(0,0) = \frac{5}{2}\frac{l(\zeta_{tail,\perp} - \zeta_{tail,\parallel})(25\zeta_{tail,\perp} + 11\zeta_{head})}{(10\zeta_{tail,\parallel} + \zeta_{head})(10\zeta_{tail,\perp} + \zeta_{head})}.$$

Finally, we obtain

$$\mathrm{curl}(g_L, g_R)(0,0) = -\frac{5}{2}\frac{l(\zeta_{tail,\perp} - \zeta_{tail,\parallel})(50\zeta_{tail,\perp} + 11\zeta_{head})}{(10\zeta_{tail,\parallel} + \zeta_{head})(10\zeta_{tail,\perp} + \zeta_{head})}.$$

| | |
|---|---|
| $M_s$ | $8 \cdot 10^5 \, Am^{-1}$ |
| $E$ | $10^{11} \, Nm^{-2}$ |
| $l_{head}$ | $10 \, \mu m$ |
| $l_{tail}$ | $10 \, \mu m$ |
| $w_{tail}$ | $1 \, \mu m$ |
| $t_{tail}$ | $0.1 \, \mu m$ |
| $\zeta_{head}$ | $0.05 \, Nsm^{-2}$ |
| $\zeta_{tail, \perp}$ | $12.4 \cdot 10^{-3} \, Nsm^{-2}$ |
| $\zeta_{tail, \parallel}$ | $6.2 \cdot 10^{-3} \, Nsm^{-2}$ |
| $B_x$ | $0.01 \, T$ |
| $B_y$ | $0.01 \, T$ |

Table 1: Values of the parameters used in the numerical simulations

Figure 1: The magneto-elastic swimmer: initial configuration, before the application of the external magnetic field. Lengths are in $\mu m$.

Figure 2: Horizontal displacement during one period of the external field (left) and velocity of the swimmer (right). Very small and very high frequencies are not effective and a maximum displacement is obtained for a frequency of about 8 Hz. Three bullets indicate the frequencies 3, 8 and 50 Hz that are used in the sequel for a more thorough analysis

Figure 3: The Fourier mode $\widehat{\Psi}_1(s)$ corresponding to the three frequencies 50 $Hz$ (black), 8 $Hz$ (red), and 3 $Hz$ (blue). The circles, represented in the complex plane, correspond to the data obtained from the numerical simulations, and we have interpolated them with straight lines. This linear approximation leads to formula (12)

Figure 4: Trajectory of the head of the swimmer with the magnetic field given by (11) with $\omega$ = 8 $Hz$. The close-up view in the right panel emphasizes the oscillations in the head movement. Lengths are in $\mu m$.

Figure 5: Trajectory of the head of the swimmer with the magnetic field given by equations (13), (14). The average direction of the magnetic field experiences a low frequency circular motion together with a high frequency oscillation. The swimmer follows the slow modulations of the applied magnetic field by tracing a circular trajectory. Lengths are in $\mu m$.

Figure 6: Trajectory of the head of the swimmer with the magnetic field given by equations (13), (15), where we have used $\omega/2\pi$ = 8 $Hz$ and $T_{max}$ = 10$s$. The sudden rotation of the axis along which the magnetic field oscillates induces a sudden change in the swimming direction that could allow the swimmer to navigate along the elbow of a pipe (not shown). Lengths are in $\mu m$.

Figure 7: Left - The loop in the $(\theta_L, \theta_R)$ traced by the MML swimmer is indicated in blue. The curve described by $(\theta_L, \theta_2)$ is given in red. Right - Snapshots of the swimming stroke along the beat cycle corresponding to points A, B, C and D.

Figure 8: Left - The loop in the $(\theta_L, \theta_R)$ plane traced by the 2-link swimmer is indicated in blue. The dashed curve gives the corresponding picture when no passive head is attached to the first link. Right - Snapshots of the swimming stroke along the beat cycle corresponding to points A, B, C and D.

Figure 9: Horizontal displacement during one cycle of the external field (left) and velocity (right) for the swimmer with a tail made of a magnetic multilayer (thickness of layers constant along filament length).

Figure 10: Horizontal displacement during one cycle of the external field (left) and velocity (right) for the swimmer with a tail made of a magnetic multilayer with an elastic joint in the middle.

Figure 1

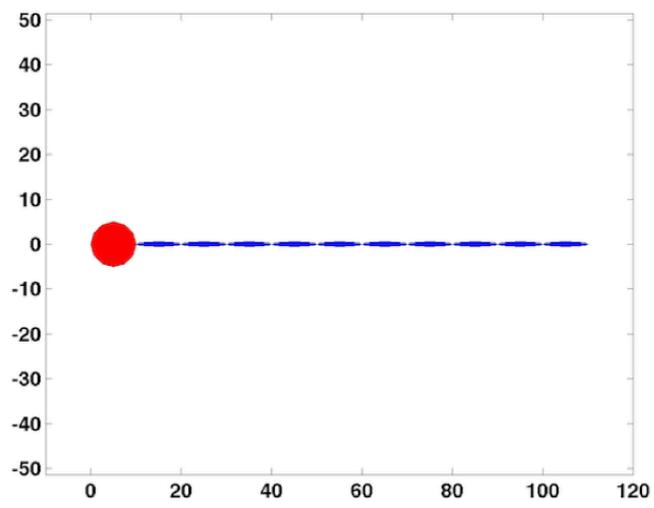

Figure 2

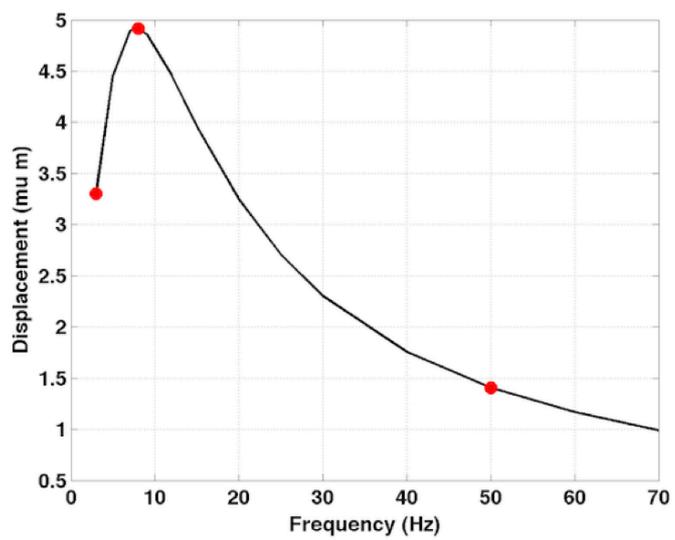 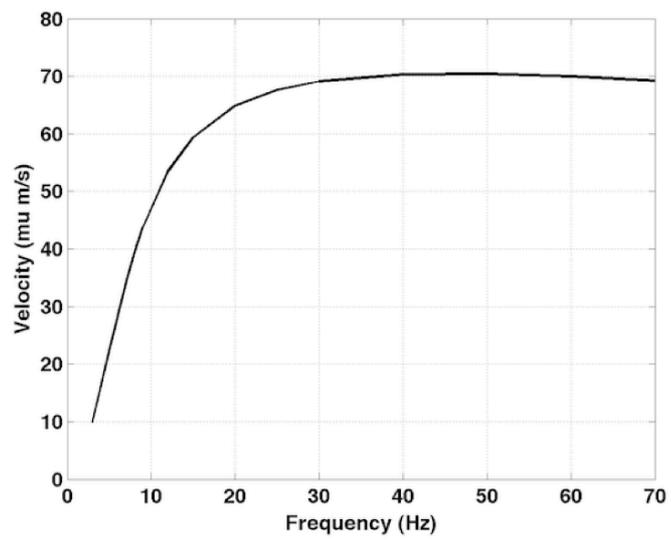

Figure 3

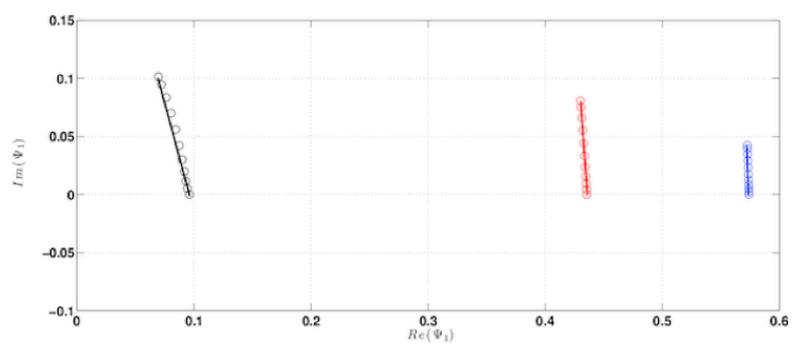

Figure 4

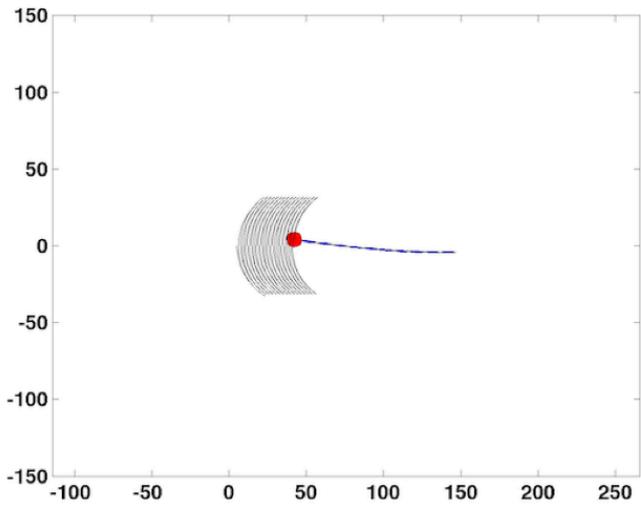 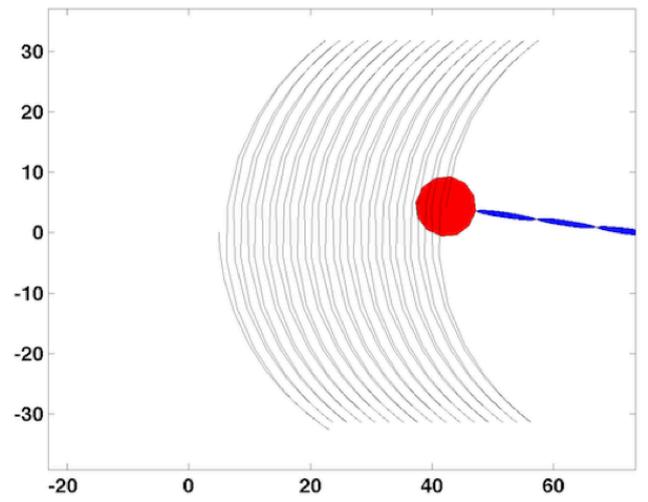

Figure 5

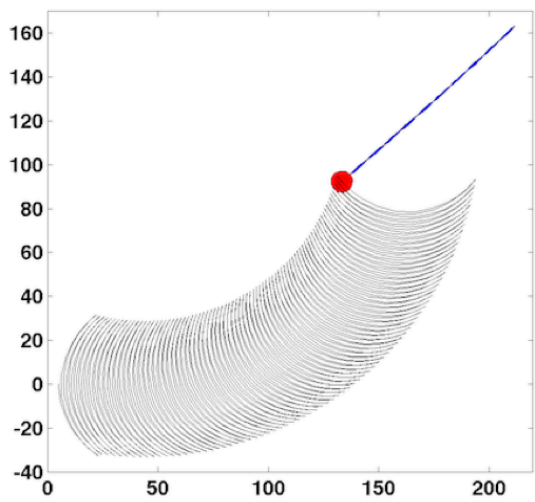

Figure 6

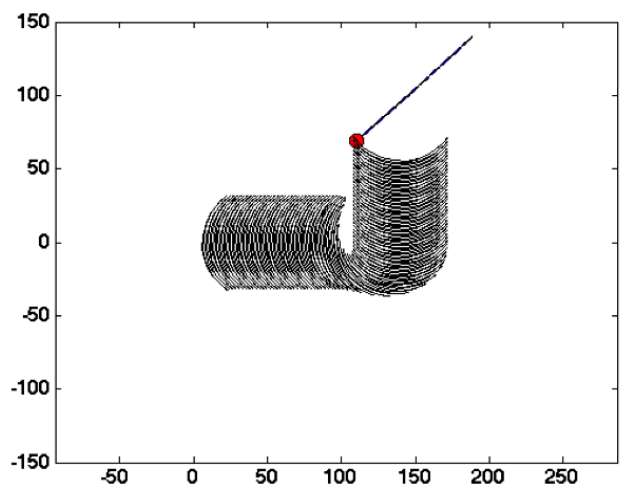

Figure 7

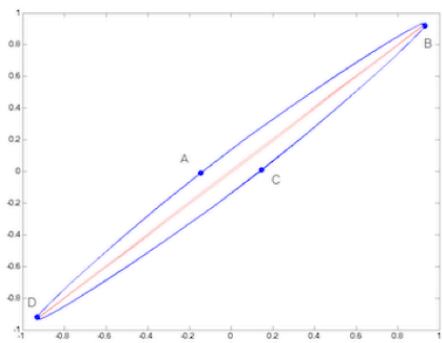 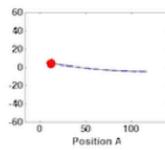 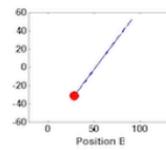
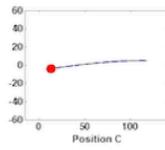 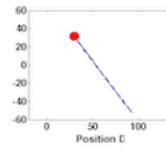

Figure 8

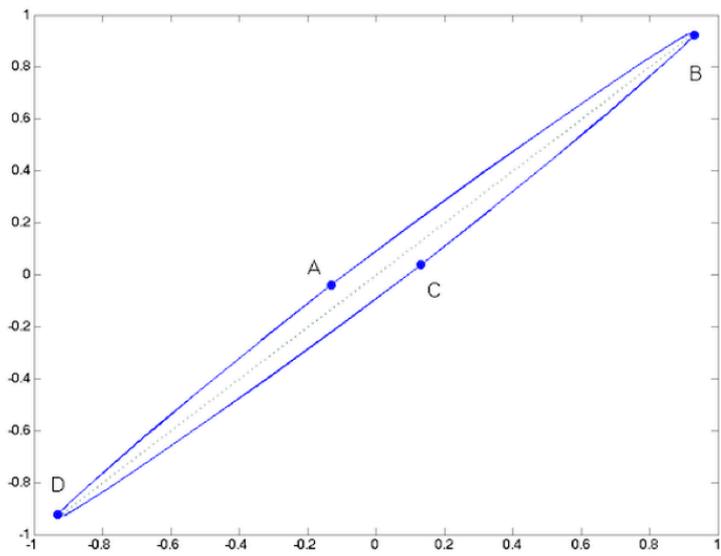
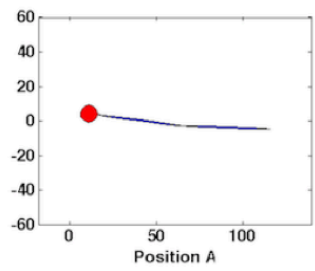
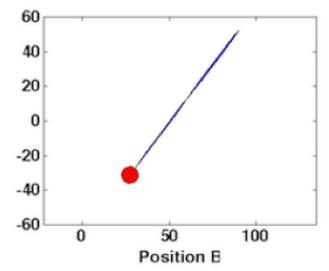
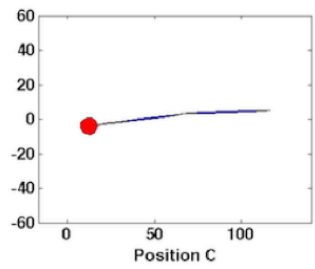
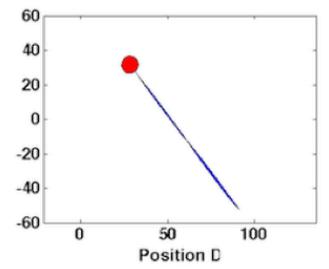

Figure 9

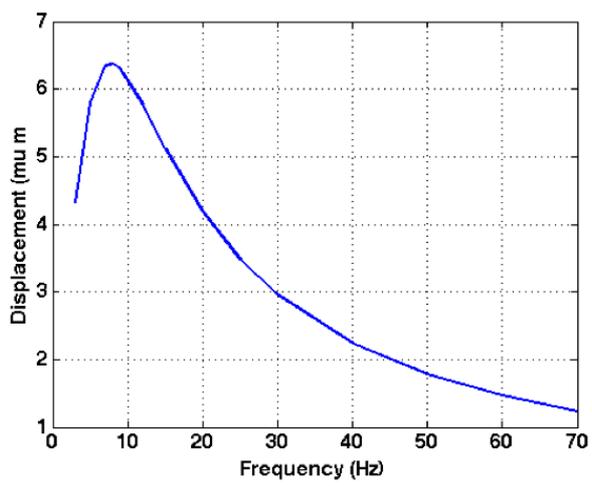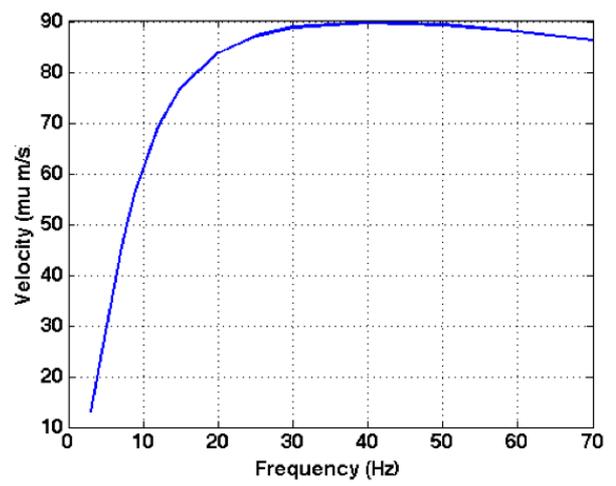

Figure 10

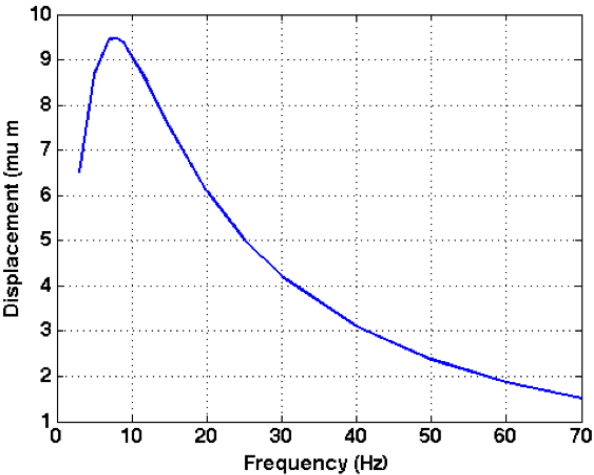 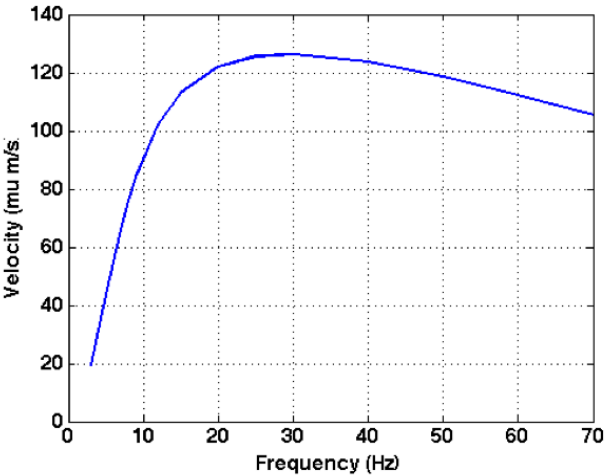